\documentclass{article}

\usepackage{PRIMEarxiv}
\usepackage[numbers,sort&compress]{natbib}
\usepackage[utf8]{inputenc} %
\usepackage[T1]{fontenc}    %
\usepackage[english]{babel}
\usepackage{etoolbox}

\let\originalselectlanguage\selectlanguage
\renewcommand{\selectlanguage}[1]{%
  \ifstrequal{#1}{en}
    {\originalselectlanguage{english}}
    {\originalselectlanguage{#1}}%
}
\usepackage[
  colorlinks=true,
  citecolor=blue,
  urlcolor=blue,
  linkcolor=blue
]{hyperref}      %
\usepackage{url}            %
\usepackage{booktabs}       %
\usepackage{amsfonts}       %
\usepackage{nicefrac}       %
\usepackage{microtype}      %
\usepackage{lipsum}
\usepackage{fancyhdr}       %
\usepackage{graphicx}       %
\usepackage{amsmath}
\usepackage{titlesec}
\usepackage{subcaption} %
\graphicspath{{figures/}}     %
\usepackage[final]{changes}
\title{Pulsed Optical Injection Steering in Multistable Semiconductor Laser Arrays under Correlated Noise

}

\author{
  Max M. Chumley\thanks{Contact Author},~~~Herbert Winful \\
  Department of Electrical Engineering and Computer Science \\
  University of Michigan \\
  Ann Arbor, Michigan, USA\\
  \texttt{mchumley@umich.edu, arrays@umich.edu} \\
}

\begin{document}
\maketitle

\begin{abstract}
We demonstrate robust programmable state preparation in small Vertical Cavity Surface Emitting Laser
(VCSEL) arrays with optical feedback using transient optical injection in the form of Gaussian pulses. In Lang-Kobayashi-type models of delay-coupled two- and three-laser arrays, multistability gives rise to coexisting synchronized and symmetry-broken equilibrium branches. We show that short injection pulses applied to one or more lasers can steer the system from free-running operation to any stable equilibrium branch in the absence of noise by appropriate choice of pulse detuning and amplitude, after which the selected state persists without continued forcing. With correlated noise, injection steering remains effective, but branches with small basins of attraction are effectively destabilized by noise. These results validate pulsed injection as a practical mechanism for attractor selection in multistable VCSEL arrays and point to a feasible route toward experimental realization of programmable collective-state control.
\end{abstract}

\section{Introduction}  

Optically coupled semiconductor laser arrays allow for increased optical power and provide control over the resulting beam properties~\cite{serkland_two-element_1999,lehman_two-dimensional_2007,xun_nineteen-element_2019,dave_static_2019}. However, due to the finite round-trip propagation time associated with optical coupling, these systems exhibit delayed dynamics that result in multistability and destabilization of synchronized states~\cite{mulet_modeling_2002,erzgraber_compound_2006,yanchuk_dynamics_2006,clerkin_multistabilities_2014,prasad_complicated_2003}. Without input into the system, depending on the parameters and the initialization, the system will either reach a stable equilibrium, approach a periodic orbit or the dynamics could be more complicated such as chaotic.  Control of VCSEL arrays has been studied in the literature through injection locking, current-tuned supermode selection, and coherent beam control~\cite{mercier_injection-locking_1996,fishman_injection-locking_1999,lucke_autostable_2000,hergenhan_fast_2000,lehman_two-dimensional_2007,thompson_mode_2019,jahan_supermode_2022,dave_static_2019}, but the use of transient optical injection to prepare selected stable branches in multistable delay-coupled arrays remains much less explored. One of the main advantages of transient optical injection is that it does not rely on a constant forcing of the system to retain its state. Prior work has established pulse-driven switching and optical memory in single bistable semiconductor lasers and VCSELs~\cite{mori_low-switching-energy_2006,brandonisio_bistability_2012}, as well as injection-induced locking and multistable synchronized states in small coupled laser arrays~\cite{mercier_injection-locking_1996,erzgraber_locking_2009,yanchuk_dynamics_2006,clerkin_multistabilities_2014}. This work extends these directions by showing that transient optical injection can be used as a state-preparation mechanism for small VCSEL arrays with optical feedback, starting from free-running lasers and steering the system toward desired stable synchronized equilibrium branches. After the injection pulse is removed, the array remains on the selected branch through its own autonomous delayed dynamics. Furthermore, the robustness of such an approach has yet to be studied with the effect of correlated noise due to photon carrier interactions and whether an array of multiple lasers can be steered to equilibrium compound laser modes (CLMs). In this paper, we study the properties of pulsed injection steering in VCSEL arrays of two and three lasers to show the robustness of this approach. The paper is organized as follows. We start by showing the model and parameters used in this work along with details on the initialization in Sec.~\ref{sec:theory}. Then, in Sec.~\ref{sec:results}, we present results examining the injection pulse shape parameters, robustness to injection-frequency errors, sensitivity to the injection phase, and performance in the presence of noise, followed by a discussion of experimental considerations. Concluding remarks are provided in Section~\ref{sec:conclusion}.

\section{Theory}\label{sec:theory}

\begin{figure}[htbp]
    \centering
    \includegraphics[width=0.6\linewidth]{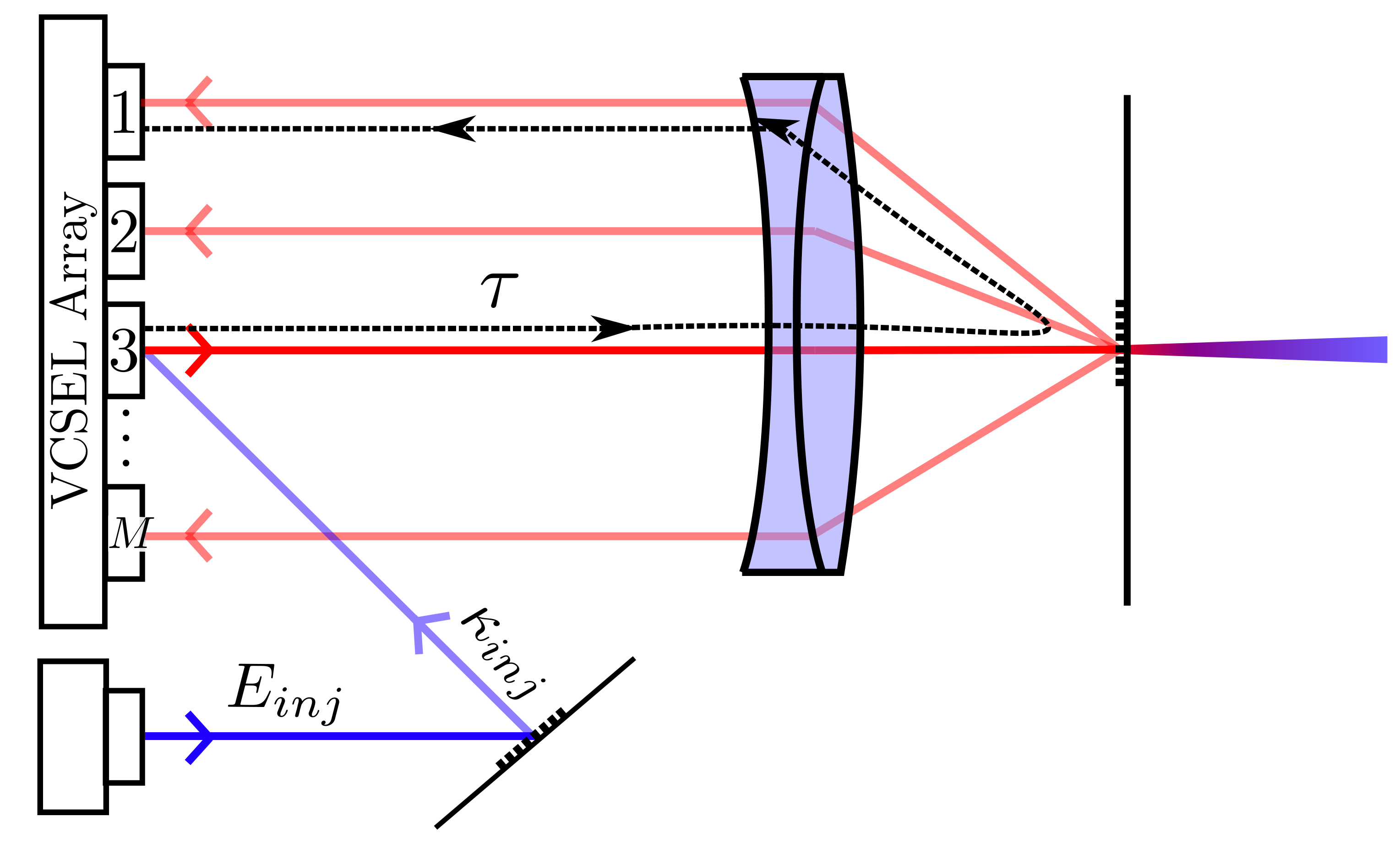}
    \caption{Injection steering schematic diagram showing the mutually coupled VCSEL array and the injection laser pulse steering the array to the frequency of the injection laser.}
    \label{fig:inj_steering_schematic}
\end{figure}

A schematic diagram for our setup is shown in Fig.~\ref{fig:inj_steering_schematic} showing the optically coupled array of $M$ lasers with a single injection pulse into laser 3 in this case. The light travels from the laser cavity to a modulator where it is partially reflected into the other lasers in the array. In principle, any laser can be coupled to any other laser in the array with arbitrary coupling topology. In Fig.~\ref{fig:inj_steering_schematic}, we show the case where laser 3 is coupled to all of the other lasers. Similarly, the injection pulse is partially reflected using a modulator and this pulse can be injected into any combination of the lasers. 
The round trip delay time $\tau$, shown as the dashed black curve, arises due to the physical distance the light traverses to reach the other laser cavities. 

\subsection{Mathematical model}

\begin{figure}[t]
    \centering
    \includegraphics[width=\textwidth]{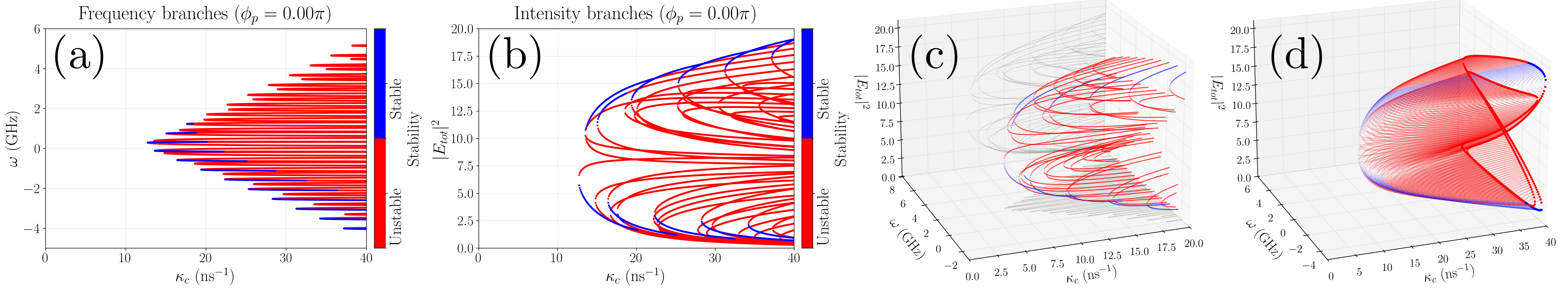}
    \caption{Equilibrium and stability branches with respect to coupling strength for a two-laser array detuned by 4 GHz. Panel (a) shows a projection of the branches on the equilibrium frequency axis, panel (b) shows the projection onto the intensity axis, panel (c) shows the 3D structure and topology of the branches with projections to show how panels (a) and (b) are related for $\phi_p=0$, and panel (d) shows the full stability manifold when $\phi_p\in[0,2\pi]$. The cross section curve is shown in bold at $\kappa_c=40$ ns$^{-1}$ to highlight the shape of the manifold.}
    \label{fig:eq_branches}
\end{figure}

We model these laser arrays using a slightly modified form of the complex coupled Lang-Kobayashi equations \cite{lang_external_1980} with correlated noise due to the photon carrier interactions \cite{fatadin_numerical_2006, ma_linewidth_2019}. $E_i(t)$ is the complex electric field of laser $i$ and $N_i(t)$ is the carrier number of laser $i$. The lasers are optically coupled using the coupling strength matrix $\kappa\in\mathbb{R}^{M\times M}$ with
\begin{equation}
    \kappa_{ij} = \kappa_c A_{ij}, 
\end{equation}
where $A$ is the adjacency matrix defining the coupling topology of the array with $\sum_{ij}A_{ij} = 1$ and $\kappa_c$ is the total coupling strength/budget. For this work, we assume $\kappa_{ii}=0$ (negligible self-feedback), $A=A^T$ (symmetric coupling), and equal coupling phases $\phi_p=\omega_0\tau$ for each coupling relationship to reduce the number of simulations but in general these parameters can vary. We introduce a time-dependent injection term $\kappa_\mathrm{inj}(t)E_{\mathrm{inj}}(t)$ to describe the optical injection steering process. This model is adapted from Ref.~\cite{ma_linewidth_2019} and mathematically represented as the system of complex stochastic delay differential equations (SDDEs),
\begin{equation}\label{eq:lk_model}
\begin{aligned}
\frac{dN_i}{dt} &=
\frac{\eta I}{q}
-\frac{N_i}{\tau_n}
-g_0\frac{N_i-N_0}{1+\gamma |E_i|^2}|E_i|^2
+F_{N_i}(t),
\\[6pt]
\frac{dE_i}{dt}
&=
\Bigg[
\frac{1}{2}
\left(
g_0\frac{N_i-N_0}{1+\gamma |E_i|^2}
-\frac{1}{\tau_p}
+\frac{\beta N_i}{\tau_n |E_i|^2}
\right)
+i\left(
\frac{\alpha}{2}g_0\frac{N_i-\bar N}{1+\gamma |E_i|^2}
+\delta_i
\right)
\Bigg]E_i
\\
&\quad
+\sum_{\substack{j=1}}^M
\kappa_{ij}E_j(t-\tau)e^{-i\phi_p}
+\kappa_{\mathrm{inj}}(t)E_{\mathrm{inj}}^i(t)
+F_{E_i}(t),
\end{aligned}
\end{equation}
where $E_\mathrm{inj}^i(t)=\sqrt{S_\mathrm{inj}}e^{i(\omega_\mathrm{inj}t+\phi_\mathrm{inj})}$ and $\kappa_\mathrm{inj}(t)=\kappa_\mathrm{inj}\exp{\left(-\frac{(t-t_p)^2}{2\tau_{w}^2}\right)}$ is the Gaussian injection pulse profile. The injection profile parameters correspond to the peak injection rate $\kappa_\mathrm{inj}$, the peak time $t_p$, and width $w_\mathrm{inj}$. We quantify the injection profile parameters as multiples of the other model parameters. For example, $\kappa_\mathrm{inj}$ is measured as a multiple of the coupling strength $\kappa_c$ (assuming all-to-all equal coupling). The remaining model parameters are defined in Table~\ref{tab:dim_params}. In this work, we restrict $E_\mathrm{inj}^i(t)=0~\forall i\neq \lfloor M/2 \rfloor$ so we only apply injection pulses to the center-most laser in the array, and for even $M$ we take the lower frequency laser of the two central lasers. This choice was driven by our goal to make this process experimentally feasible, but in principle the pulses can be applied to any combination of the lasers. The model follows a slightly modified Lang–Kobayashi formulation, where the gain term is referenced to the transparency carrier density $N_0$, while the phase dynamics are referenced to the free-running steady-state carrier density $\bar{N}$, consistent with prior linewidth and frequency-noise analyses \cite{ma_linewidth_2019}. Due to the drastically different time-scales present in the system between the carrier and photon lifetimes, the system in Eq.~\ref{eq:lk_model} is inherently stiff.
\added{For numerical convenience, we convert the equations to a dimensionless form by scaling time by $\tau_p$. This normalization improves conditioning and simplifies the implementation, while preserving the underlying dynamics. This transformation is defined in detail in Appendix~\ref{app:dim_model}. All simulations shown in this article use the dimensionless model, however, all of our results are converted back to their dimensional form using the inverse transformation for convenient interpretation. }
\begin{table}
\centering
\caption{Model parameters and dimensional values}
\setlength{\tabcolsep}{6pt}
\renewcommand{\arraystretch}{1.2}

\resizebox{\textwidth}{!}{%
\begin{tabular}{|c|c||c|c|}
\hline
Injection current ($I$) & $3I_{\mathrm{th}}$ 
& Photon lifetime ($\tau_p$) & $5.4$ ps \\ \hline

Threshold current ($I_{\mathrm{th}}$) 
& $\frac{q}{\tau_n}\left(N_0+\frac{1}{g_0\tau_p}\right)$ 
& Differential gain ($g_0$) & $8.75\times10^{-4}$ ns$^{-1}$ \\ \hline

Carrier injection rate ($\eta$) & 0.9 
& Linewidth enhancement factor ($\alpha$) & 2 \\ \hline

Electron charge ($q$) & $1.602\times 10^{-19}$ C 
& Transparency carrier number ($N_0$) & $2.86\times10^5$ \\ \hline

Carrier lifetime ($\tau_n$) & $0.25$ ns 
& Free running carrier number ($\bar{N}$) & $5.13\times10^5$ \\ \hline

Gain saturation ($\gamma$) & $4\times10^{-6}$ 
& Spontaneous emission factor ($\beta$) & $1\times10^{-3}$ \\ \hline

$i$th detuning frequency ($\delta_i$) & $\in[0,5]$ GHz 
& Feedback delay ($\tau$) & $1$ ns \\ \hline

Total coupling strength ($\kappa_c$) & $\in[0,40]$ ns$^{-1}$ 
& Coupling phase matrix ($\phi_p$) & $\in[0,2\pi]$ \\ \hline
\end{tabular}%
}

\label{tab:dim_params}
\end{table}
Equilibrium points and their stability for this model were computed using the approach in Appendix~\ref{app:stability_analysis} over a range of coupling strengths ($\kappa_c$) and coupling phase ($\phi_p$) to demonstrate the strong multistability present in the system. We plotted equilibrium branches for an array of two lasers detuned by 4 GHz in Fig.~\ref{fig:eq_branches} where, panel (a) shows the CLM frequency branches, panel (b) shows the total field intensity branches, and panel (c) shows how these branches are related in three-dimensional (3D) space along with how they connect through varying the coupling phase in panel (d) to form a smooth stability manifold. 

\subsection{History Function}

\begin{figure}
    \centering
    \includegraphics[width=0.9\textwidth]{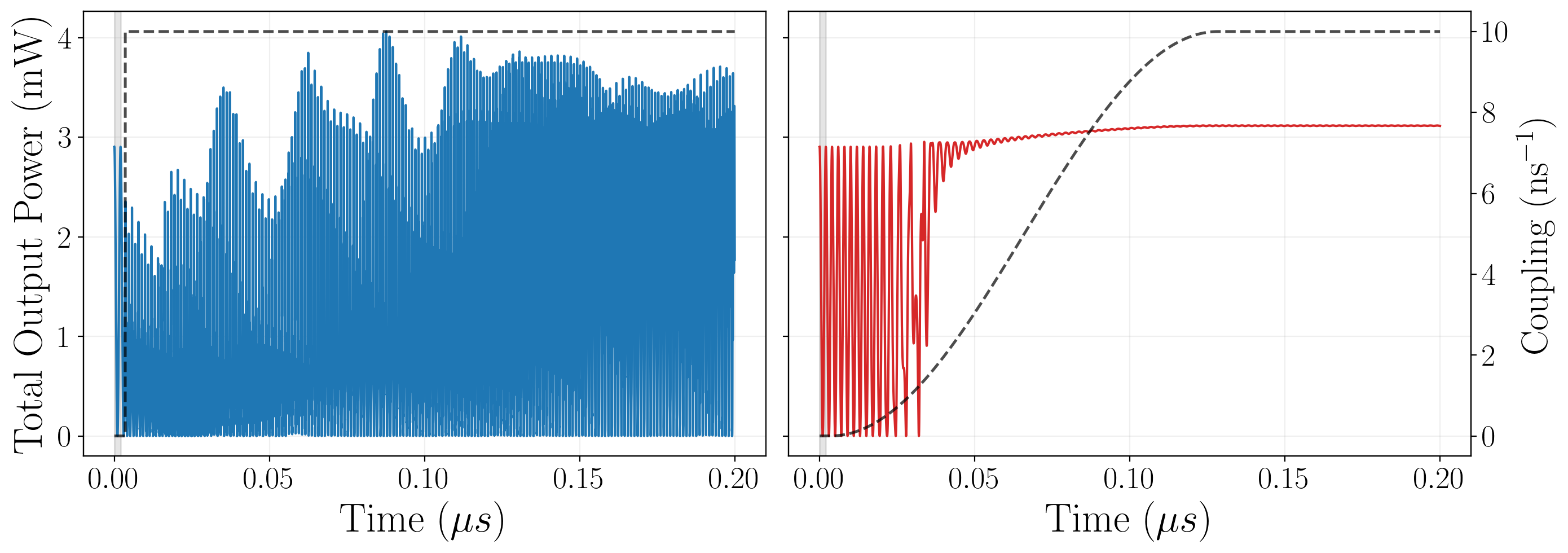}
    \caption{Example time-series plots where the coupling is increased at different rates. The left panel shows a chaotic response when the coupling ramp is sharp and the example on the right uses a gradual ramp to reach a fixed point of the system. This demonstrates that the initial discontinuities present when the history differs from a true steady state of the system can push the system to an entirely different attractor if not mitigated properly.}
    \label{fig:coupling_ramp}
\end{figure}

A critical difference between delay differential equations (DDEs) and ordinary differential equations (ODEs) is that for DDEs the choice of initial condition or history function is infinite dimensional. In order to integrate these systems using the numerical techniques outlined in Appendix~\ref{app:stability_analysis}, the states need to be defined for $t\in[-\tau,0]$. Specifically for solving these LK models, the history function is typically set to be either free-running $(\kappa=\mathbf{0})$, zero-field, where all of the lasers are inactive for $t<0$ or including noise in the initial history \cite{nair_using_2021, ye_optimal_2025}. A significant issue arises at $t=0$ in the form of an initial discontinuity depending on the choice of history. This phenomenon is usually ignored when studying steady-state behavior of a DDE as it typically does not change the long term response, but as we show in Fig.~\ref{fig:eq_branches}, these systems are highly multistable so a discontinuous jump in any of the states at $t=0$ can quickly lead to nonphysical behavior by pushing the trajectory to a different stable branch. Our work avoids this problem by starting all simulations at the free running solution (unless otherwise specified) and slowly ramping the coupling from 0 to $\kappa_c$. We argue that this results in a smooth transition between the history and steady-state behavior and more accurately models how these systems are studied in experimental settings. The ramp profile we chose is the cosine ramp,
\begin{equation}
    \kappa_{ij} (t) = \frac{\kappa_cA_{ij}}{2} \left(1-\cos{\left(0.8\pi\frac{(t-t_s)}{t_r}\right)}\right),
\end{equation}
where $t_s$ is the start time of the ramp and $t_r$ is the rise time to transition from 10\% to 90\% of the final value. If $t_r\to 0$, this essentially becomes a step function at $t_s$ resulting in discontinuous behavior but allowing the coupling strength to gradually increase over many delay cycles significantly reduces this effect. This is demonstrated in Fig.~\ref{fig:coupling_ramp}, where the total output power for two lasers is shown for two different coupling ramp profiles. 
While ultimately the choice of history does not have any effect on the system behavior with injection steering, we chose to apply a ramp to the coupling to our simulations the avoid the artificial discontinuities that come with simulating DDEs and provide a smooth transition between the free-running and coupled dynamics. 

\section{Results}\label{sec:results}

\begin{figure}[t]
    \centering
    \includegraphics[width=\textwidth]{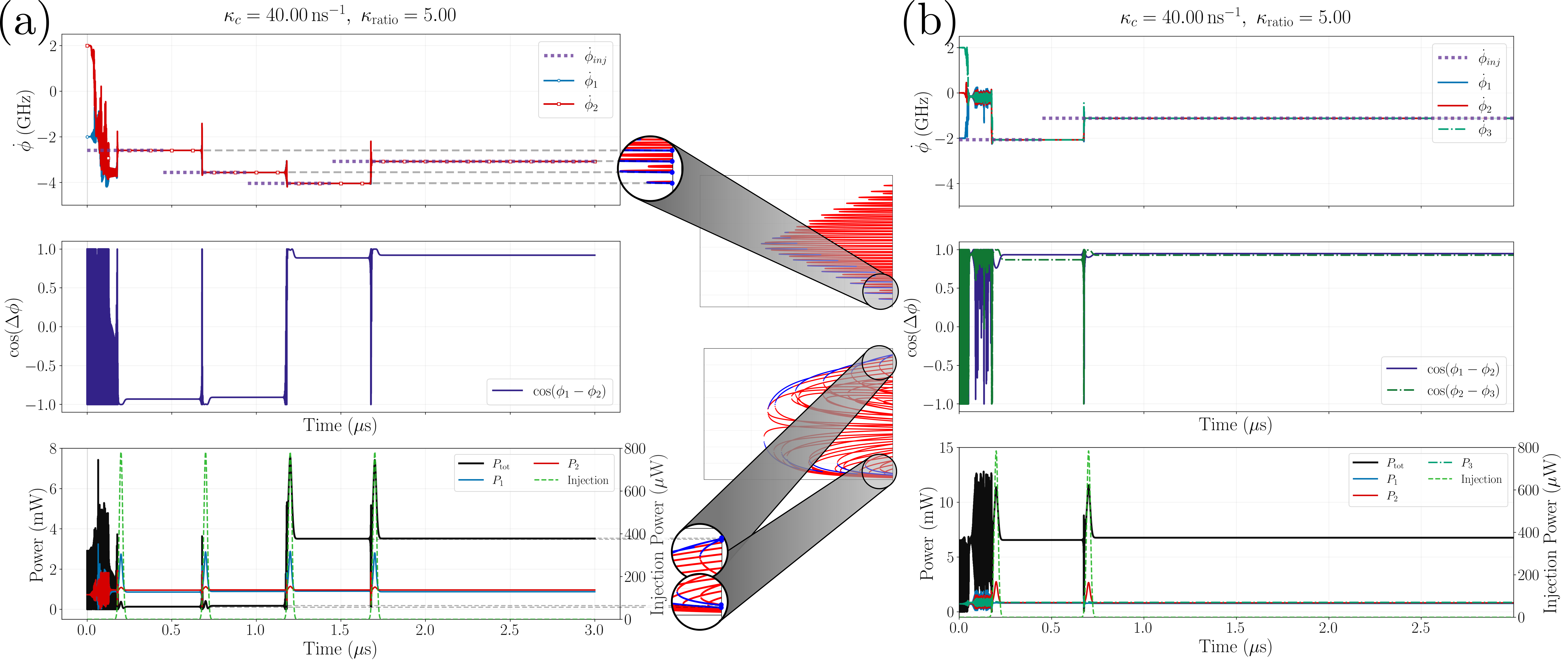}
    \caption{\added{Injection steering examples for two [panel (a)] and three laser [panel (b)] arrays to demonstrate the process of starting with an array of free running lasers and steering the array to all of the stable equilibria of the system using controlled injection pulses. The top rows show the lasing frequencies of each laser and the dashed lines show the equilibrium frequencies for each injection. In the middle rows, we plotted the cosine of the phase difference, and in the bottom rows we plotted the coupling/injection power and the output power for the coupled beam. The dashed curves indicate the mutual coupling ramp, and the injection steering pulses and their magnitudes are indicated on the right-hand axis in microwatts. We also show for the two-laser case how these frequencies and output powers relate to the equilibrium branches in the middle.}}
    \label{fig:injection_steering_TD}
\end{figure}

To test the robustness and capabilities of injection steering, we started by applying it to an array of two coupled lasers. We set the total coupling to $\kappa_c=40~$ ns$^{-1}$ and the detuning frequency vector was set to $\vec{\delta}=[-2,2]$ GHz so that the total detuning is 4 GHz. Stable equilibrium solutions were obtained and sorted by intensity from lowest to highest using the approaches in Appendix~\ref{app:stability_analysis}. Four stable solutions were obtained for the laser array. The system was simulated for 3 $\mathrm{\mu s}$ starting with a free running history and $\kappa_c$ was varied from 0 to 40 ns$^{-1}$ starting after ten delay cycles using the cosine ramp profile with a rise time of $50\tau$. After $0.2~\mathrm{\mu s}$, an injection pulse was applied to the first equilibrium solution with $\tau_w=10\tau$ and $\kappa_\mathrm{inj}=3\kappa_c$ and consecutive pulses were applied every $0.5~\mathrm{\mu s}$ to steer the system to the other stable equilibrium points. This process is shown in Fig.~\ref{fig:injection_steering_TD} for two and three lasers with the same amount of total coupling for comparison. The top panels show the lasing frequencies and the equilibrium frequencies used for injection steering. In the middle panels, the cosine of the phase differences are plotted to show the phase properties of the equilibrium points and in the bottom panels the output power is plotted over time along with the injection power. We see that for the same total coupling, the two-laser system has more stable equilibrium points compared to only two in the three-laser system. For the two-laser case we also show that the steering can go in both directions by steering back down to the out-of-phase solutions in the same simulation. 
To determine the injection power, we assumed the injection laser was just a simple free running semiconductor laser with a wavelength of $\lambda=910$ nm. Using the peak injection strength and steady-state photon number $\bar{S}$, the injection power was computed as
\begin{equation}
    P_\mathrm{inj} = \frac{hc}{\lambda}\kappa_\mathrm{inj}\bar{S}. 
\end{equation}
To get the output power of the laser array, a similar approach was used but with the photon decay rate $1/\tau_p$ setting the emission timescale instead of $\kappa_\mathrm{inj}$. This yields an expression for the output power of each laser of the form
\begin{equation}
P_{\mathrm{out}} = \frac{hc}{\lambda}\frac{\bar{S}_{\text{tot}}}{\tau_p},
\end{equation}
which corresponds to the rate at which photons leave the cavity. The total output power of the array is then obtained by using the total photon number $\bar{S}_{\text{tot}}$. In Fig.~\ref{fig:injection_steering_TD}, we see that the injection power is on the order of hundreds of microwatts while the output power is on the milliwatt scale so the output power is significantly higher than the injection steering pulse power required to steer the system to the desired equilibrium point. 

\subsection{Injection strength and width testing}\label{sec:inj_strength_width}

The injection steering strength profile is given by the Gaussian pulse distribution,
\begin{equation*}
    \kappa_\mathrm{inj}(t)=\kappa_\mathrm{inj}\exp{\left(-\frac{(t-t_p)^2}{2\tau_{w}^2}\right)}.
\end{equation*}
In order to understand the limits of injection steering, we needed to test the injection profile parameter dependence and define bounds for the parameters to yield a successful injection steering pulse. We remark that injection topology is another property that can be considered, but we chose to only apply this injection to the centermost laser to aid in experimental feasibility. First, we consider the peak injection strength $\kappa_\mathrm{inj}$ and injection width $\tau_w$ for testing. To measure the success of the injection steering pulse, we define the locking parameter,
\begin{equation*}
    L = |\dot{\phi}(t) - \omega_\mathrm{inj}|,
\end{equation*}
so after the injection pulse a successful outcome at steady state is indicated by the average, \added{$\mu(L)\to 0$ and the standard deviation $\sigma(L)\to 0$. In other words, the lasing frequency remains at the desired frequency after the pulse and the standard deviation is close to zero indicating steady-state or fixed-point behavior. To test the robustness of different parameters, we compute $\mu(L)$ and $\sigma(L)$ for various coupling strengths and detuning frequencies and average the results using}
\begin{align}
    \langle\mu(L)\rangle&=\frac{1}{N_\kappa N_\delta}\sum_{i=1}^{N_\kappa}\sum_{j=1}^{N_\delta}\mu\bigl(L(\kappa_i,\delta_j)\bigr),\\
    \langle\sigma(L)\rangle&=\frac{1}{N_\kappa N_\delta}\sum_{i=1}^{N_\kappa}\sum_{j=1}^{N_\delta}\sigma\bigl(L(\kappa_i,\delta_j)\bigr),
\end{align}
\added{where $\langle\cdot\rangle$ denotes averaging over all $(\kappa_f,\delta)$ parameter combinations. Throughout the article and figure titles, “Average $\mu(L)$” denotes the quantity $\langle\mu(L)\rangle$, i.e., the average of $\mu(L)$ over all sampled $(\kappa_f,\delta)$ values.}

\begin{figure}[htbp]
    \centering
    \includegraphics[width=0.9\linewidth]{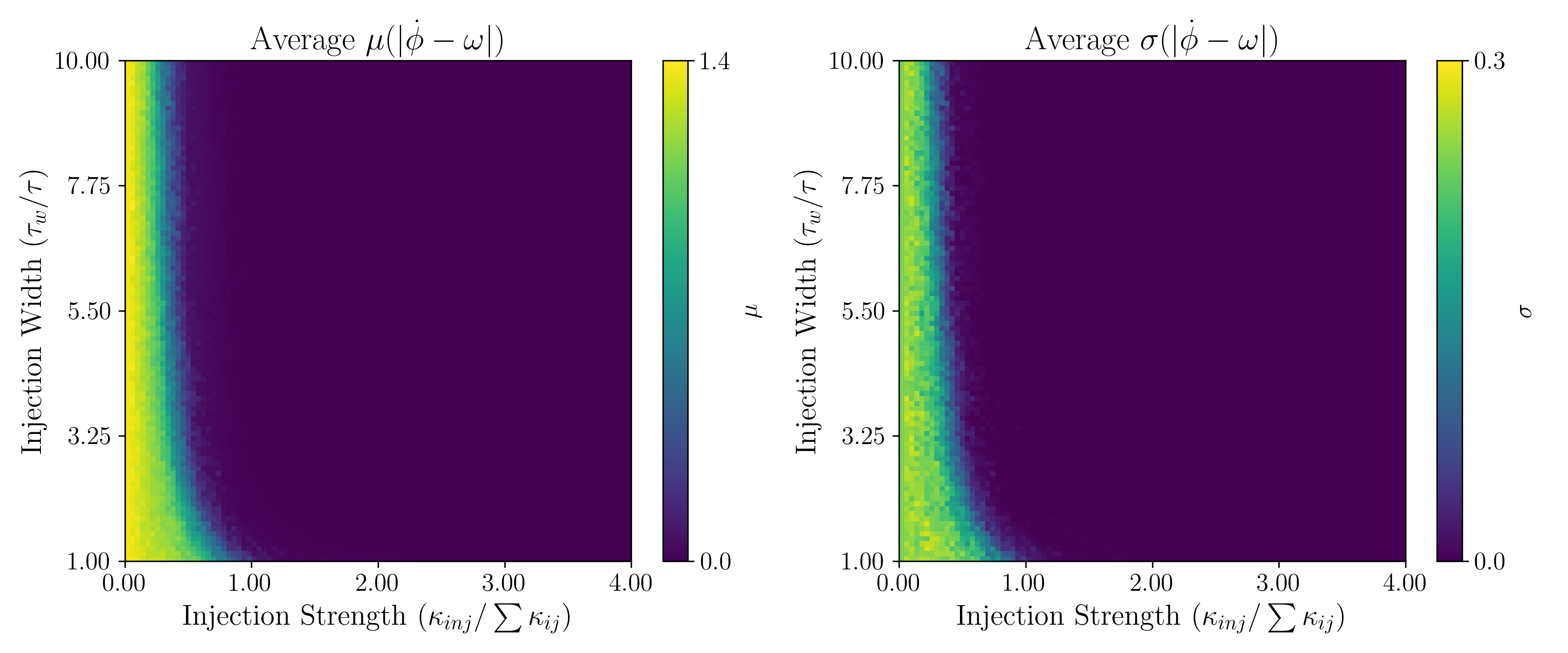}
    \caption{Test results varying the injection steering pulse strength and width over many different coupling strengths and detuning frequencies for a two-laser array. }
    \label{fig:inj_shape_2laser}
\end{figure}

\begin{figure}[htbp]
    \centering
    \includegraphics[width=0.9\textwidth]{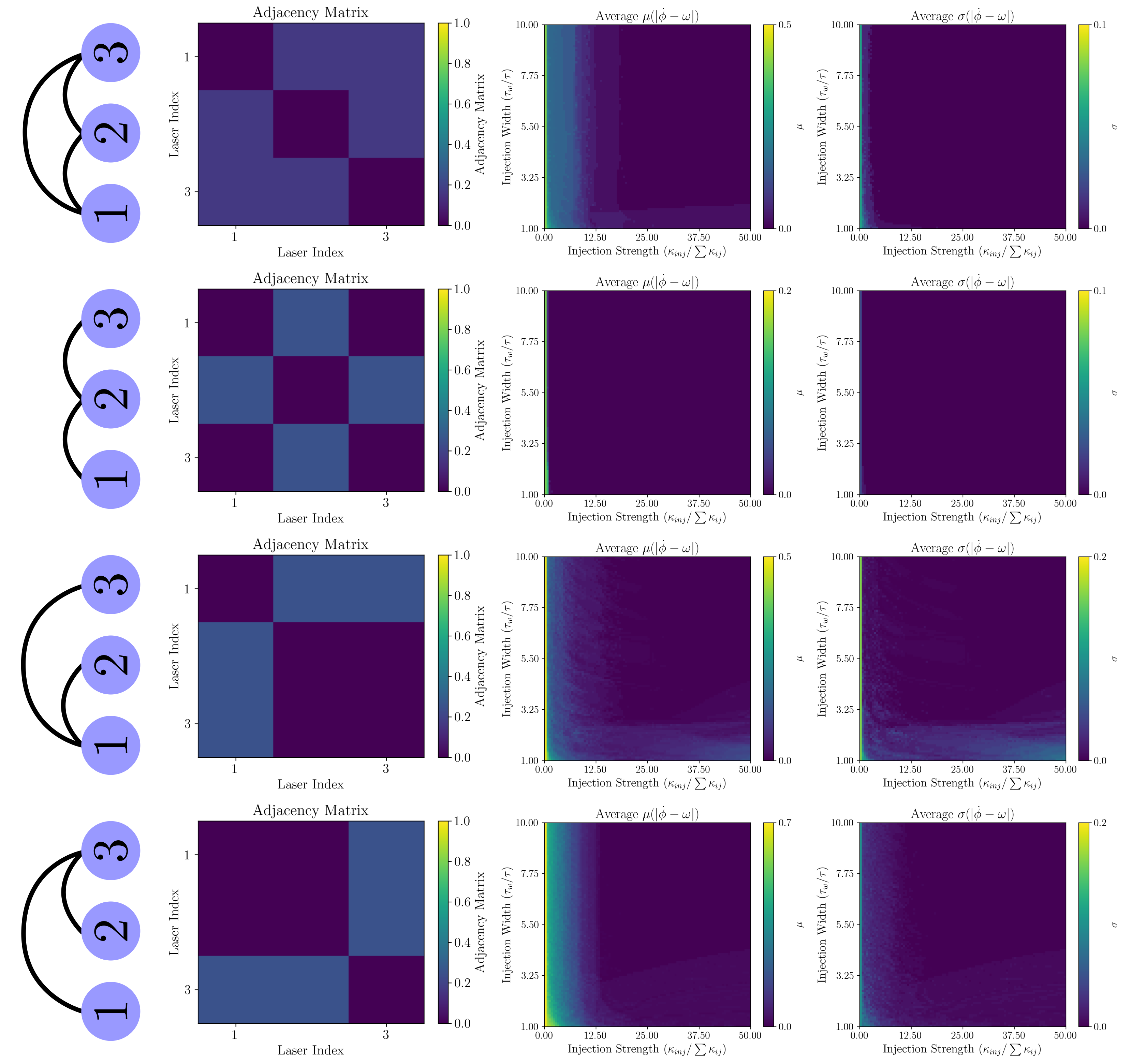}
    \caption{Test results varying the injection steering pulse strength and width over many different coupling strengths and detuning frequencies for a three laser array. }
    \label{fig:inj_shape_3laser}
\end{figure}

We simulated a two-laser array from free running history for 0.5~$\mathrm{\mu s}$ ramping the coupling from 0 to $\kappa_f$ with a detuning frequency $\delta$ between the lasers. An injection pulse was introduced at $t_p=0.2~\mathrm{\mu s}$ using pulse parameters over $(\kappa_\mathrm{inj},\tau_w)\in [0,4\kappa_c]\times[\tau, 10\tau]$ and the final $0.3 \mathrm{\mu s}$ were used to compute $L$. This process was repeated for $\kappa_f\in [0.001,40]~\mathrm{n s}^{-1}$ and $\delta\in[0,5]~$GHz to average the results over many configurations and determine the required injection strength and width. The final parameter to consider is the injection phase $\phi_\mathrm{inj}$. Due to the difficulty of controlling the phase in an experiment, we assume that $\phi_\mathrm{inj}\sim\mathcal{U}(-\pi,\pi)$ to measure success of injection steering independent of the injection phase. The results for two lasers are shown in Fig.~\ref{fig:inj_shape_2laser}. 
We see that on average $\mu(L)$ and $\sigma(L)$ are consistently near zero for $\kappa_\mathrm{inj}>\kappa_c$ and $\tau_w>3\tau$. The same tests were also conducted for the three-laser system with all four unique coupling topologies as shown in Fig.~\ref{fig:inj_shape_3laser}. 
We see that larger injection strengths are required for the three laser array because only one laser recieves the pulse and more power is required to influence the entire array.

\subsection{Injection frequency and phase testing}

\begin{figure}[htbp]
    \centering
    \includegraphics[width=0.9\linewidth]{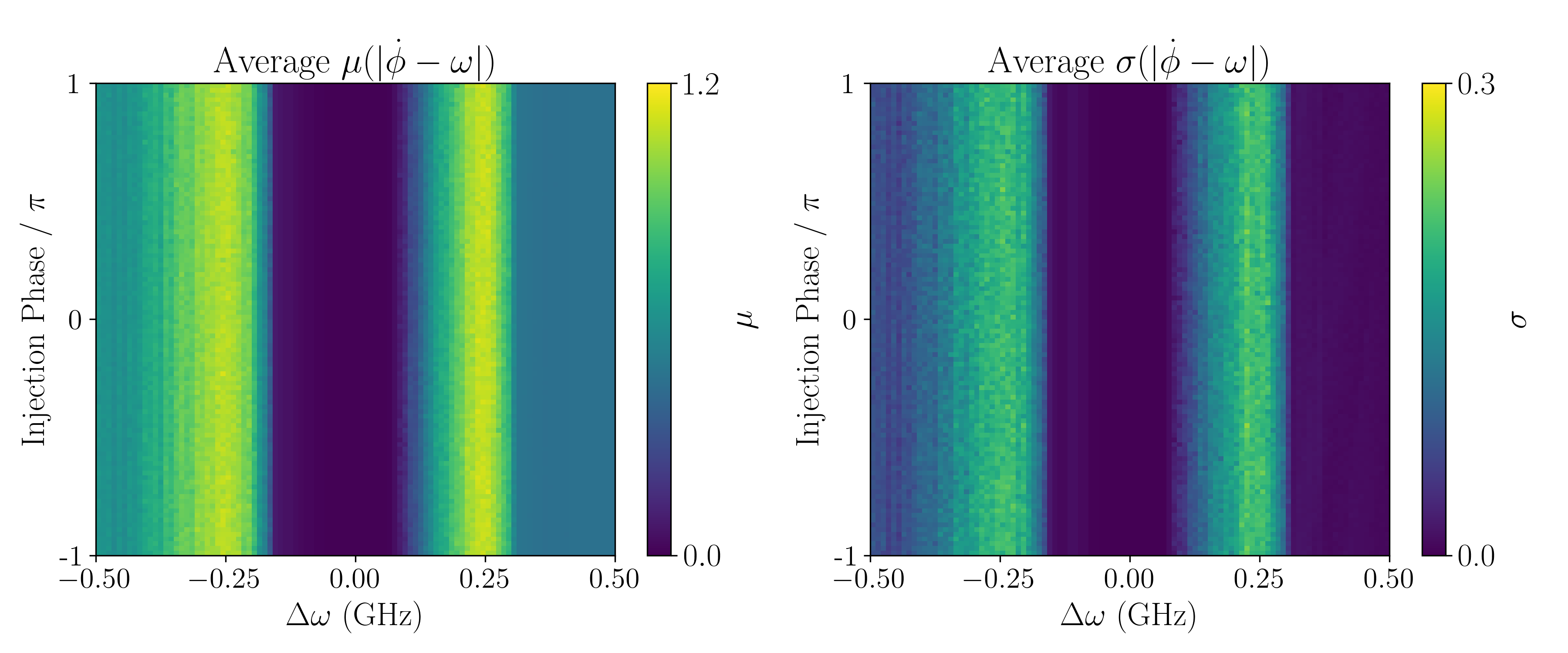}
    \caption{Test results varying the injection steering pulse frequency and phase over many different coupling strengths and detuning frequencies for a two-laser array. }
    \label{fig:inj_freq_phase_2laser}
\end{figure}

\begin{figure}[t]
\centering
\begin{minipage}{0.49\textwidth}
    \centering
    \includegraphics[width=\linewidth]{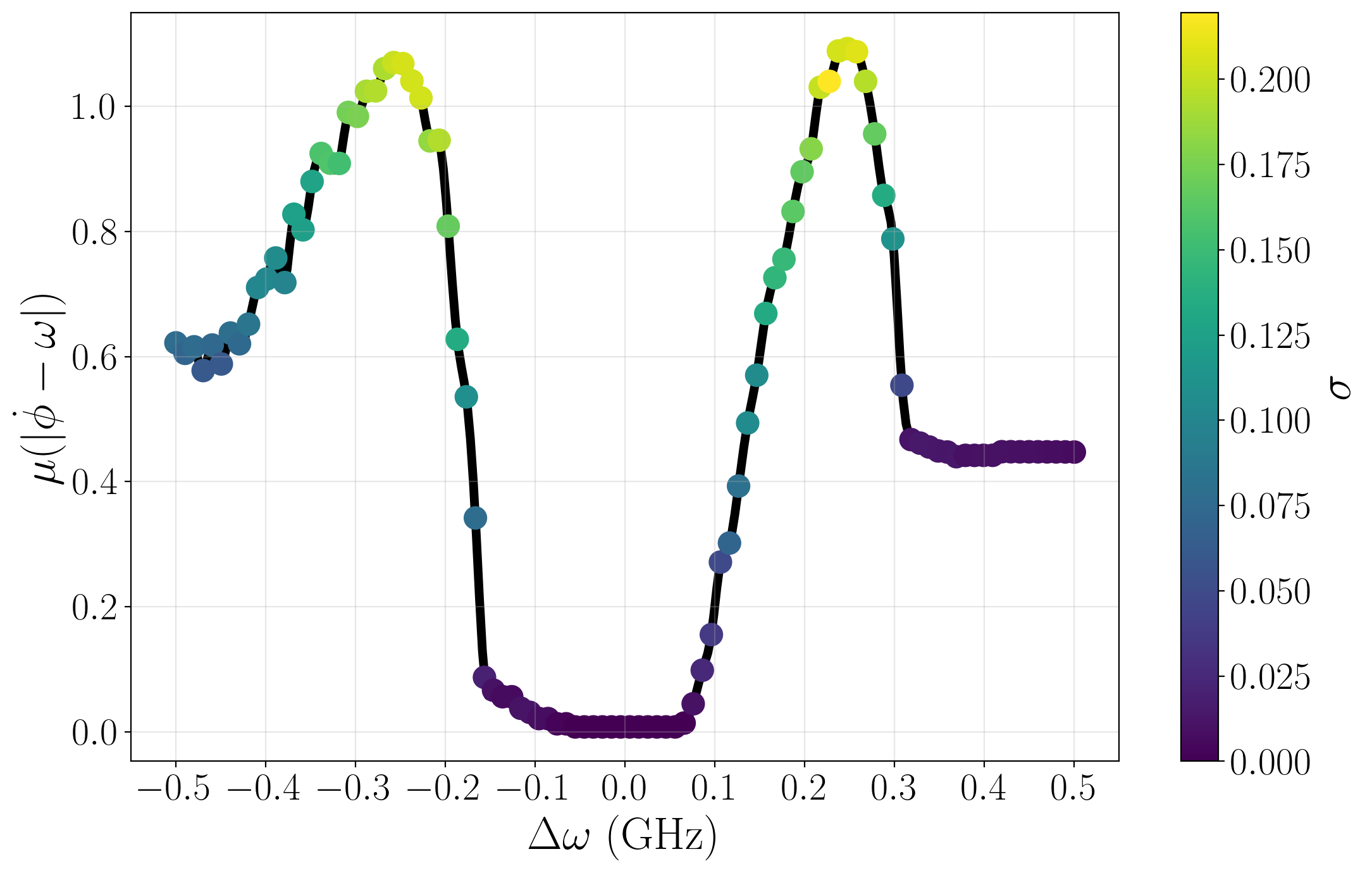}
    {\small (a)}
\end{minipage}
\hfill
\begin{minipage}{0.49\textwidth}
    \centering
    \includegraphics[width=\linewidth]{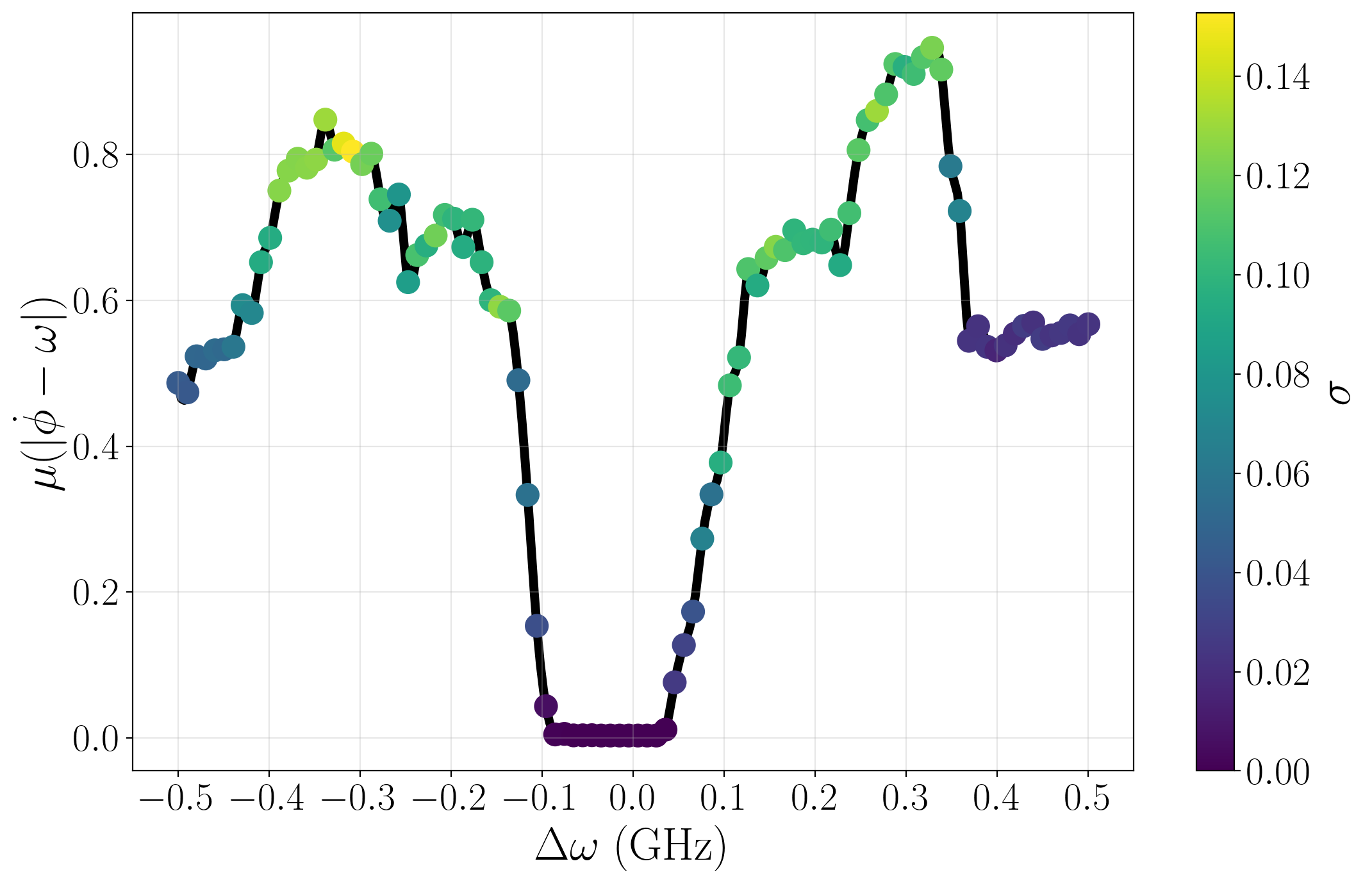}
    {\small (b)}
\end{minipage}
\caption{Injection frequency offset test results compressed into two dimensions by averaging over the injection phase axis. The color indicates the standard deviation of $L$. (a) Two-laser array and (b) three-laser array.}
\label{fig:2Dinj_freq_phase}
\end{figure}

Next, we fixed $\kappa_\mathrm{inj}=5\kappa_c$ and $\tau_w=5\tau$ and test the injection frequency $\omega_\mathrm{inj}$ and phase $\phi_\mathrm{inj}$. For this injection strength, this corresponds to a peak injection power of approximately $235~\mathrm{\mu W}$ at the largest coupling strength tested. In previous results, $\omega_\mathrm{inj}$ was identically equal to the equilibrium frequency $\omega_{\mathrm{eq}}$ of the system that yielded the maximum output intensity. However, this may not be possible in practice so it is important to test how close the frequency needs to be for injection steering to work. We also test the injection phase here to further verify the independence of the injection phase on the final outcome. For this test, we defined $\omega_\mathrm{inj}=\omega_{\mathrm{eq}}\pm \Delta \omega$ for some $\Delta \omega\in[-0.5,0.5]~$ GHz and $\phi_\mathrm{inj}\in[-\pi,\pi]$ and simulated the system with the same setup as Sec.~\ref{sec:inj_strength_width} computing the average and standard deviation $L$ at each pair of $(\Delta \omega, \phi_\mathrm{inj})$. The results for this test are shown in Fig.~\ref{fig:inj_freq_phase_2laser}, where we see further proof of the independence of $\phi_\mathrm{inj}$ and a range of $\Delta\omega$ near zero between approximately --0.15 and 0.1 GHz. 
This result was also compressed into a single two-dimensions (2D) plot due to the independence of $\phi_\mathrm{inj}$ as shown in Fig.~\ref{fig:2Dinj_freq_phase}(a) and the same result was generated for a three-laser array with all-to-all coupling.
This result shows that if $|\Delta\omega|$ is too large, it is possible for the injection steering to push the system to another stable equilibrium point that is clear for $\Delta\omega>0.3$ GHz, where the average $L$ is around 0.4 GHz and the standard deviation is close to zero. 

\subsection{Noise robustness}
When simulating this system without noise, the injection steering process was shown to successfully navigate to the desired equilibrium point on average over a wide range of injection pulse shapes and frequencies. To further mimic an experimental procedure, we introduce the correlated noise terms in Eq.~\eqref{eq:lk_model} and study the average behavior over many iterations. \added{We first study the noise robustness with respect to the injection frequency offset ($\Delta \omega$), and then we discuss the effects of higher spontaneous emission percentage ($\beta$), which increases the noise amplitude. }

\subsubsection{Frequency offset robustness}

\begin{figure}[t]
\centering
\begin{minipage}{0.49\textwidth}
    \centering
    \includegraphics[width=\linewidth]{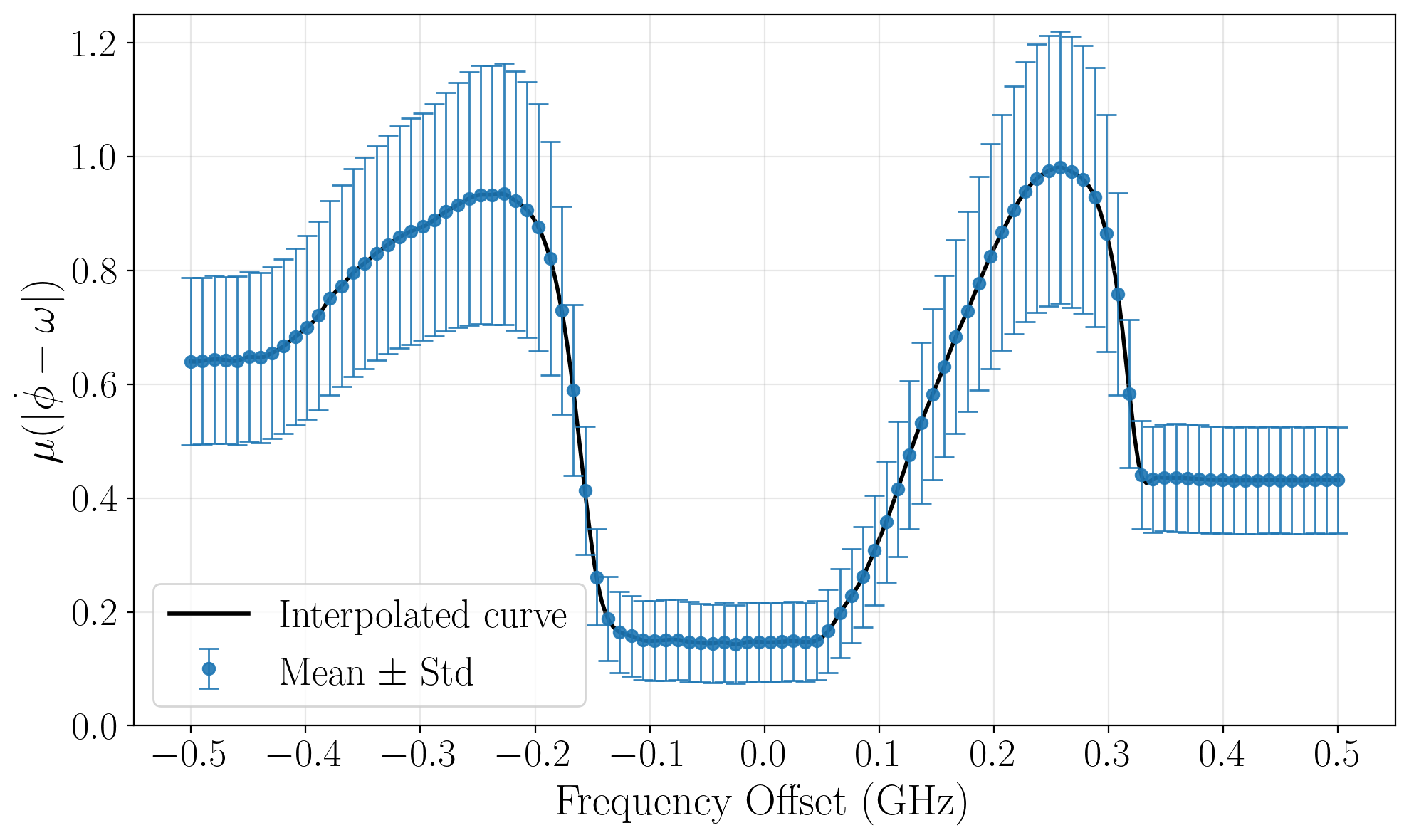}
    {\small (a)}
\end{minipage}
\hfill
\begin{minipage}{0.49\textwidth}
    \centering
    \includegraphics[width=\linewidth]{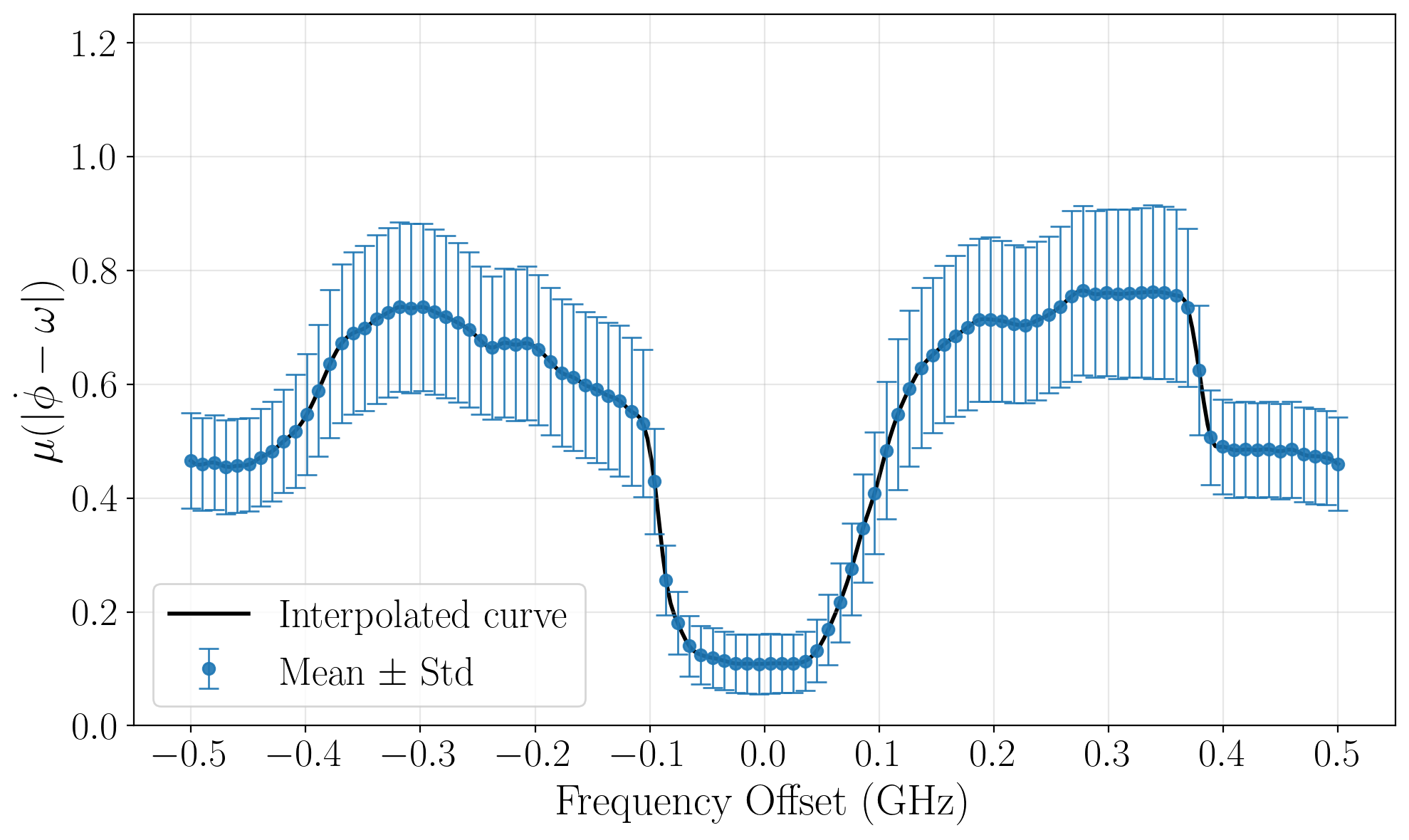}
    {\small (b)}
\end{minipage}
\caption{Noise robustness test over 100 noise iterations to measure the mean and standard deviation $L$ with noise. (a) Two-laser array and (b) three-laser array.}
\label{fig:noise_robustness}
\end{figure}

\begin{figure}[htbp]
    \centering
    \includegraphics[width=0.9\linewidth]{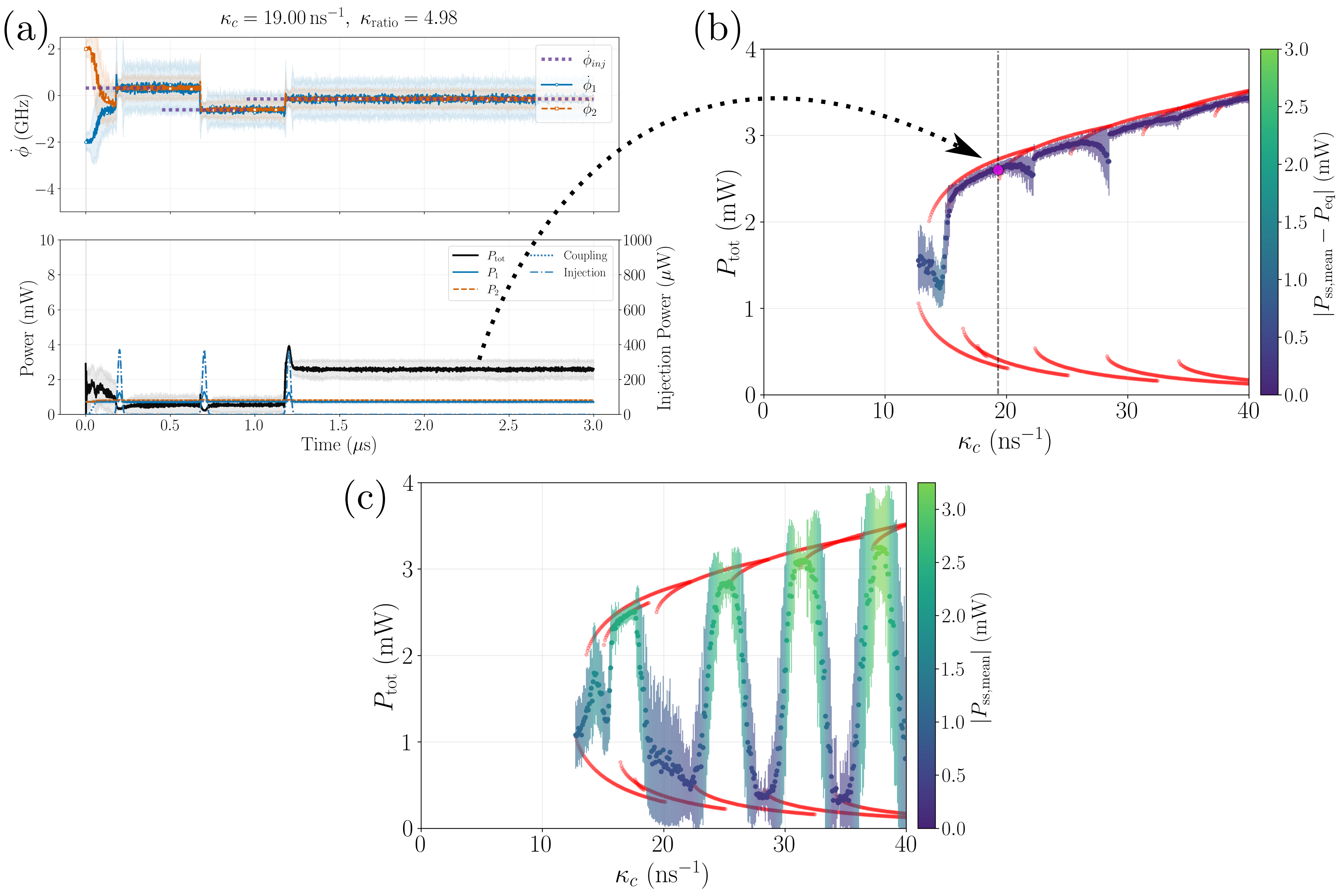}
    \caption{\added{Injection steering results with noise. Panel (a) shows a single example applying injection steering to a two-laser system with a detuning of 4 GHz at a coupling strength of 19 ns$^{-1}$. At these parameters, the noise-free system has three stable equilibrium points and applying injection pulses the system is successfully steered to all three branches. Panel (b) shows the stable equilibrium branches in black and the steady-state mean of the trajectories in blue with corresponding standard deviation error bars. In this case, only the extrema of the equilibrium with the highest intensity are shown for clarity. Panel (c) demonstrates the effectiveness of injection steering by plotting the percentage difference compared to the target equilibrium output power and panel (d) shows the stationary distribution mean and standard deviation of the same system without any injection pulses to show the effectiveness of this approach.}}
    \label{fig:noise_stability_manifold}
\end{figure}

For each frequency offset, we compute the average and standard deviation $L$ for 100 noise iterations with randomly chosen injection phases. The results for this test are shown in Fig.~\ref{fig:noise_robustness} for both two- and three-laser arrays. It is clear that the noise robustness testing has the same general shape as the plot without noise and the valley with the smallest mean is where the injection steering is successful on average near $\Delta\omega=0$.  
While Fig.~\ref{fig:noise_robustness} shows that injection steering is successful on average with noise included, we decided to fix the detuning at 4 GHz and visualize how the stability manifold changes when noise is included. In Fig.~\ref{fig:eq_branches}(b), we see the stability manifold for this system with the blue branches indicating a stable equilibrium point at a particular coupling strength. When noise is introduced into the system, one would expect a slightly smaller output power/intensity because the noise causes the phase to vary more. However, Fig.~\ref{fig:eq_branches}(b) cannot be directly created for the system with noise because the stationary distribution needs to be obtained through simulation. To quantify the stability manifold of the system with noise, we simulated the system with 100 noise realizations at every coupling strength and applied injection steering to all of the stable equilibria found at each point with the pulses separated by 0.5 $\mathrm{\mu s}$ as shown in Fig.~\ref{fig:noise_stability_manifold}(a) for $\kappa_c=19$ ns$^{-1}$. At this coupling strength, the system has three stable equilibrium points and we see in the stationary distribution time series that on average the injection steering successfully navigated the system to all three points. Performing this process for every coupling strength, but only steering to the equilibrium with highest intensity and recording the average and standard deviation after the injections, we obtained the manifold shown in Fig.~\ref{fig:noise_stability_manifold}(b). \added{The black points indicate the stable equilibrium points without noise and the color of the remaining points indicates the percentage difference from the desired equilibrium point after injection. In other words, blue points with small error bars suggest that the trajectory stabilized very close to the target equilibrium point. }
It is clear that the in-phase equilibrium points persist when noise is included, but their output powers differ due to the imperfect synchronization from the relatively high noise amplitudes. Furthermore, gaps form between the equilibrium branches due to the noise where the top equilibrium points destabilize for a small range of coupling strengths as we can see in Fig.~\ref{fig:noise_stability_manifold}(b) where the branches start to drop around 22 and 28 ns$^{-1}$. The power of injection steering is further highlighted when the stationary distributions are computed without any injection pulses as shown in Fig.~\ref{fig:noise_stability_manifold}(d). The stationary distribution appears to switch between in-phase (blue) and out-of phase (red) as the coupling strength is increased and even when the points are near the true equilibrium, oscillations are still present in the signals. Whereas in Fig.~\ref{fig:noise_stability_manifold}(b), the maximum intensity solution can be chosen for nearly every coupling strength in this range and the standard deviation is significantly lower. \added{The percent differences are shown on the same plot in Fig.~\ref{fig:noise_stability_manifold}(c) with and without injection. We quantify a successful injection steering case as a percent difference of less than 5\%, which is indicated by the green box on Fig.~\ref{fig:noise_stability_manifold}(c). It is clear that significantly more points are inside of the green region when injection steering is enabled compared to without injection. }

\subsubsection{Spontaneous Emission Dependence}

\added{To test the noise robustness, the spontaneous emission factor ($\beta$) must also be tested due to its direct impact on the amplitude of the noise in this model. Researchers have found that for VCSELs, $\beta \in [10^{-5},10^{-3}]$ \cite{yang_green_2023} but can be larger in some cases \cite{liu_importance_2015}. Throughout this work, we have assumed $\beta=10^{-3}$ and have shown that injection steering is quite reliable for this case. In this section, we tested injection steering for $\beta \in [10^{-5},10^{-2}]$ to determine how much noise is required to result in unsuccessful steering. The results for these different spontaneous emission factors are shown in Fig.~\ref{fig:beta_dep} where we see successful steering cases for $\beta \in [10^{-5},10^{-3}]$, but when too much spontaneous emission is present ($\beta=10^{-2}$) injection steering fails to steer the system to the desired equilibrium points. }
\begin{figure}[htbp]
    \centering
    \includegraphics[width=\textwidth]{beta_dependence}
    \caption{\added{Spontaneous emission factor ($\beta$) dependence on injection steering success. $\beta$ is varied from $10^{-5}$ to $10^{-2}$ to test the robustness of injection steering to the noise level.}}
    \label{fig:beta_dep}
\end{figure}

\subsection{Experimental considerations}
To this point, the base assumption for this method is that the equilibrium frequency or destination for the system is known prior to injection. In an experiment, however, this is often initially unknown. For this method to be beneficial in an experiment, an approach needs to be provided that can help experimenters use injection steering to synchronize laser arrays. In experiments with delay-coupled semiconductor lasers, a scan of the detuning between lasers is often achieved by varying the temperature or the pump current of one of the lasers \cite{erzgraber_mutually_2005}. Monitoring the intensity steps as the lasers lock into a common frequency makes it possible to map the stable compound laser modes. Once the equilibrium frequencies are determined, the output of a master laser can be frequency shifted and directed as an injection pulse into any of the coupled lasers by using an acousto-optic modulator.

\section{Conclusion}\label{sec:conclusion}

A comprehensive study was performed applying pulsed optical injection steering in semiconductor laser arrays where short injection pulses were used to steer the system to a known stable equilibrium point. Injection pulse shape parameters were tested to validate the requirements for the strength of the pulse and its duration. Tests were also conducted to determine the dependence on the phase of the injection pulse and the frequency offset from the true equilibrium frequency. It was found that on average the frequency can vary by approximately 0.1 GHz from the true equilibrium and the results are independent of the injection phase. Correlated noise was also included in the model to test the noise robustness of this method for experimental feasibility. Our results show that injection steering is successful under noise for many of the equilibrium points, but for the CLMs that are destabilized by noise due to their small basins of attraction, injection steering fails to reach those branches with noise. This method allows for controlling the output of the array with minimal injection input to the system, but requires knowledge of the equilibrium frequencies for success. The approach used for finding equilibrium points and their stability quickly breaks down computationally for larger arrays because it becomes increasingly difficult to solve the equilibrium equations and stability eigenvalue problem due to the dimensionality of the problem. In the future, we aim to leverage machine learning techniques to drive the system to a synchronized state rather than requiring prior information about the underlying equilibrium topology of the system.

\section{Acknowledgments}
This material is based upon work supported by Office of Naval Research under award number MURI ONR-N000142412548. The authors acknowledge useful discussions with Hui Cao and Nathan Vigne of Yale University.

\section{Code Availability}\label{sec:code}
We have made our code publicly available in a Github
repository in Ref.~\cite{chumley_vcsel_lib_2026}.

\appendix

\section{Dimensionless Model Derivation}\label{app:dim_model}

\subsection{Component model}

Before converting the model in Eq.~\eqref{eq:lk_model}, we first needed to split the electric field into separate amplitude and phase equations. This is done by setting $E_i(t)=\sqrt{S_i(t)}\exp{(\phi_i(t) t)}$ and separating the real and imaginary terms to obtain equations for the time derivatives of the photon number $(\dot{S}_i)$ and phase $(\dot{\phi}_i)$, where the over dot is the time derivative. The resulting equations are
\begin{equation}\label{eq:dim_comp_model}
\begin{aligned}
\frac{dN_i}{dt} &= \frac{\eta I}{q} - \frac{N_i}{\tau_n} - g_0 \frac{N_i - N_0}{1+\gamma S_i} S_i + F_{N_i},
\\[8pt]
\frac{dS_i}{dt} &=
g_0 \frac{N_i - N_0}{1+\gamma S_i} S_i
- \frac{S_i}{\tau_p}
+ \beta \frac{N_i}{\tau_n}
+ F_{S_i}
\\
&\quad
+ 2 \sum_{\substack{j=1 \\ j \neq i}}^M
\kappa_{ij}\sqrt{S_j(t-\tau)S_i(t)}
\cos\!\left(\phi_j(t-\tau)-\phi_p-\phi_i(t)\right)
\\
&\quad
+ 2\kappa_{\mathrm{inj}}(t)\sqrt{S_{\mathrm{inj}}S_i(t)}
\cos\!\left(\omega_{\mathrm{inj}} t+\phi_{\mathrm{inj}}-\phi_i(t)\right),
\\[8pt]
\frac{d\phi_i}{dt} &=
\frac{\alpha}{2} g_0 \frac{N_i-\bar{N}}{1+\gamma S_i}
+ \delta_i
+ F_{\phi_i}
\\
&\quad
+ \sum_{\substack{j=1 \\ j \neq i}}^M
\kappa_{ij}\sqrt{\frac{S_j(t-\tau)}{S_i(t)}}
\sin\!\left(\phi_j(t-\tau)-\phi_p-\phi_i(t)\right)
\\
&\quad
+ \kappa_{\mathrm{inj}}^i(t)\sqrt{\frac{S_{\mathrm{inj}}}{S_i(t)}}
\sin\!\left(\omega_{\mathrm{inj}} t+\phi_{\mathrm{inj}}-\phi_i(t)\right).
\end{aligned}
\end{equation}
It is important to point out in this step that the phase equation has an $S_i(t)$ term in the denominator, which can be dangerous if the photon number approaches zero. In order to avoid issues in the integration, the photon number was always clipped to be larger than $10^{-11}$. To simulate photon-carrier interactions, we include the correlated noise model from Refs.~\cite{fatadin_numerical_2006, ma_linewidth_2019}, which is mathematically expressed as
\begin{align*}
F_{N_i} &=
    \sqrt{\frac{2N_i(t)}{\tau_n}}
    \, x_{1i}
    \;-\;
    \sqrt{\frac{2 \beta N_i(t)S_i(t)}{\tau_n}}
    \, x_{2i}
 \,, \\
F_{S_i} &= 
    \sqrt{\frac{2 \beta N_i(t)S_i(t)}{\tau_n}}
    \, x_{2i} , \\
F_{\phi_i} &= 
    \sqrt{\frac{\beta N_i(t)}{2 \tau_n S_i(t)}}
    \, x_{3i} \, , \\
x_i&\sim \mathcal{N}(0, \Delta t).
\end{align*}
In practice, $\tau_n\gg\tau_p$ and for the results in this work, $\tau_n$ is roughly 50 times larger than $\tau_p$. These vastly different time-scales present in the system cause numerical stiffness during integration if they are not properly accounted for. To mitigate this issue, we converted the dimensional component model to a dimensionless form by scaling time by the photon lifetime. Mathematically,
\begin{equation*}
    t' = \frac{t}{\tau_p}.
\end{equation*}
This scaling leads to the transformation outlined in Table~\ref{tab:transformation} adopted from Ref.~\cite{flunkert_delay-coupled_2011}.
\begin{table}
\centering
\caption{Dimensionless model transformation \cite{flunkert_delay-coupled_2011}}
\setlength{\tabcolsep}{10pt}
\renewcommand{\arraystretch}{1.3}
\begin{tabular}{|c|c|c|c|}
\hline
Parameter & Symbol & Transformation & Dimensionless value \\ \hline

Scaled photon number & $s_i(t')$ & $g_0\tau_n S_i(t)$ & --\\ \hline
Scaled carrier number & $n_i(t')$ & $\frac{N_i(t)-(g_0\tau_p)^{-1}}{N_0+(g_0\tau_p)^{-1}}$ & --\\ \hline
Time scale ratio & $T$ & $\tau_n/\tau_p$ & 46.296\\ \hline
Scaled delay & $\tau'$ & $\tau/\tau_p$& 185.185\\ \hline
Gain saturation & $\gamma'$ & $\gamma/(g_0\tau_n)$& 0.0183\\ \hline
Coupling strength & $\kappa'$ & $\tau_p\kappa$& $[0,0.216]$\\ \hline
Coupling phase & $\phi_p$ & $\phi_p$& $[0,2\pi]$\\ 
\hline
Pump Rate & $p$ & $g_0\tau_p (I \tau_n /q - N_0) - 1$& $3.998$\\ 
\hline
Linewidth enhancement factor & $\alpha$ & $\alpha$& 4.0\\ 
\hline
Detuning frequency & $\delta'_i$ & $\delta_i\tau_p$& $[-0.085,0.085]$\\ 
\hline
Spontaneous emission factor & $\beta$ & $\beta$& $0.001$\\ 
\hline
Spontaneous emission bias & $\beta_c$ & $\beta(g_0\tau_p N_0 + 1.0)$& $0.00235$\\ 
\hline
\end{tabular}
\label{tab:transformation}
\end{table}
\begin{equation}\label{eq:nd_comp_model}
\begin{aligned}
\frac{dn_i}{dt'} &= \frac{1}{T}\left[p - n_i - \frac{1+n_i}{1+\gamma' s_i}s_i\right] + f_{n_i},
\\[8pt]
\frac{ds_i}{dt'} &=
\left(\frac{1+n_i}{1+\gamma' s_i}-1\right)s_i
+ \beta n_i + \beta_c + f_{s_i}
\\
&\quad
+ 2 \sum_{\substack{j=1 \\ j \neq i}}^M
\kappa'_{ij}\sqrt{s_j(t'-\tau')\,s_i(t')}
\cos\!\left(\phi_j(t'-\tau')-\phi_p-\phi_i(t')\right)
\\
&\quad
+ 2\kappa_{\mathrm{inj}}'(t')\sqrt{s_{\mathrm{inj}}^{\,i}\,s_i(t')}
\cos\!\left(\omega_{\mathrm{inj}}' t' + \phi_{\mathrm{inj}}^{\,i} - \phi_i(t')\right),
\\[8pt]
\frac{d\phi_i}{dt'} &=
\frac{\alpha}{2}\frac{n_i-\bar n}{1+\gamma' s_i}
+ \delta_i' + f_{\phi_i}
\\
&\quad
+ \sum_{\substack{j=1 \\ j \neq i}}^M
\kappa'_{ij}\sqrt{\frac{s_j(t'-\tau')}{s_i(t')}}
\sin\!\left(\phi_j(t'-\tau')-\phi_p-\phi_i(t')\right)
\\
&\quad
+ \kappa_{\mathrm{inj}}'(t')\sqrt{\frac{s_{\mathrm{inj}}^{\,i}}{s_i(t')}}
\sin\!\left(\omega_{\mathrm{inj}}' t' + \phi_{\mathrm{inj}}^{\,i} - \phi_i(t')\right).
\end{aligned}
\end{equation}
Finally, we used the transformations in Table~\ref{tab:transformation} to convert the noise model to a dimensionless form. We point out that the variance of the normal distribution was scaled into the $t'$ space using $\Delta t'=\Delta t/\tau_p$ to obtain the proper cancellation in deriving the dimensionless noise model. The dimensionless noise model is given by
\begin{align*}
f_{n_i} &=
    \sqrt{\frac{2 g\tau_p(n_i + n_0 + 1)}{T}}
    \, x_{1i}
    \;-\;
    \sqrt{\frac{2 \beta (n_i + n_0 + 1) s_i}{T^2}}
    \, x_{2}
 \,, \\
f_{s_i} &= 
    \sqrt{2 \beta (n_i + n_0 + 1) s_i}
    \, x_{2i} , \\
f_{\phi_i} &= 
    \sqrt{\frac{\beta (n_i + n_0 + 1)}{2 s_i}}
    \, x_{3i} \, , \\
x_i&\sim \mathcal{N}(0,\Delta t').
\end{align*}

\section{Integration Scheme and Stability Computation}
\label{app:stability_analysis}

\subsection{Integration scheme}
\added{The system in Eq.~\ref{eq:lk_model} is a system of SDDEs with correlated noise. While many software packages exist for solving ODEs, DDEs and stochastic differential equations \cite{matlab_dde23,ansmann_efficiently_2018}, these packages do not allow for integrating systems with both delay and noise terms. As a result, for this study we needed to choose a numerical integration scheme that optimizes for computational efficiency while also supporting the SDDE framework with a stiff system. Many papers using similar delay models opt for using high-order integration schemes like fourth order Adam Bashforth Moulton for integrating these stochastic equations \cite{nair_using_2021}. These predictor-corrector integration schemes impose significant computational overhead especially when simulating over many parameters or with $M\gg 2$. The computational complexity comes from the number of evaluations of the model Eq.~\eqref{eq:lk_model} where the number of equations scales as $3M$ but a single evaluation of the model has a time complexity $\mathcal{O}(M^2)$ due to the coupling. If one naively minimizes the number of evaluations of the model, explicit Eulers method would be chosen due to requiring one evaluation per time step. However, explicit Euler is known to require exceedingly small time steps for stiff dynamical systems \cite{butcher_general_2006}. A better approach is to use implicit Eulers method where the next state depends on the gradient of the system at that step. This method performs much better with stiff systems due to the L-stability \cite{butcher_general_2006}. For this work, a hybrid method was used,}
\begin{equation}
    y_{n+1}=y_n+\Delta t\left[(1-\theta)f(y_n)+\theta f(y_{n+1})\right],
\end{equation}
\added{where $\theta=0$ corresponds to explicit Euler and $\theta=1$ is implicit Euler. To balance accuracy and mitigate stiffness issues simultaneously, we set $\theta=0.5$, which corresponds to the second-order Crank-Nicholson scheme. This $\theta$ method is A-stable for $\theta \ge 0.5$, with the trapezoidal rule ($\theta=0.5$) achieving second-order accuracy but lacking L-stability, which can lead to weak damping of stiff modes \cite{hairer_numerical_nodate}. The noise was included using the Euler-Maruyama updates in the form}
\begin{equation}
    y_{n+1} \leftarrow y_{n+1} + G(y_n),
\end{equation}
\added{where $G(y_n)$ is a sample from the noise model in Appendix~\ref{app:dim_model}.}

\subsection{Equilibrium computation}\label{sec:eq_comp}

To find equilibrium points for this system, we follow a procedure similar to that in Ref.~\cite{barioni_interpretable_2025}. We define a residual vector function $\mathbf{r}(\mathbf{x}): \mathbb{R}^{2M}\to\mathbb{R}^{2M}$ as
\[
\mathbf{r}
=
\big[
\dot s_1,\;\ldots,\;\dot s_M,\;
\dot{\phi}_{12},\;\ldots,\;\dot{\phi}_{1M},\;
\dot{\phi}_1 - \omega
\big].
\]
where $\dot{\phi}_{i1}(t)=\dot{\phi}_i(t)-\dot{\phi}_1(t)$ is the time varying lasing frequency difference between laser $i$ and laser $1$, and $\omega$ is the common equilibrium frequency. We note that  the quantity $\phi_j(t-\tau)-\phi_i(t)$ appears in the model and at the locked frequency $\omega$, $\phi_j(t-\tau)=\omega (t-\tau)+\phi_j^*$ and $\phi_i(t)=\omega t+\phi_i^*$. Combining these definitions, we get the relationship, $\phi_j(t-\tau)-\phi_i(t)=\tilde{\phi}_{ij}-\omega\tau$ where $\tilde{\phi}_{ij}\equiv \phi_j^* - \phi_i^*$ with $\phi_1^*=0$ is the steady-state phase offset between laser $i$ and laser $j$. Now, we simultaneously solve the following nonlinear system of equations
\begin{equation}
    \mathbf{r}(\mathbf{x}^*)=\mathbf{0}
\end{equation}
for the optimal steady state $\mathbf{x}^*\in\mathbb{R}^{2M}=\left[s_1^*,\ldots,s_M^*,\phi_{2}^*,\dots,\phi_{M}^*,\omega\right]$. Notably, we do not include the equilibrium carrier densities in $\mathbf{x}^*$ because using Eq.~\eqref{eq:nd_comp_model}, $n^*_i$ can be computed from $s^*_i$ using
\begin{equation}
    n^*_i = \left(1+\frac{\added{s^*_i}}{1+\gamma' s^*_i}\right)^{-1}\left(p-\frac{s^*_i}{1+\gamma' s^*_i}\right).
\end{equation}
For this work, we used the Powell hybrid method in the SciPy optimization library \cite{scipy_hybr_root} to find the equilibrium points. We remark that this approach requires an initial guess and the algorithm converges to a solution from that guess. To find as many equilibrium solutions as possible, we provided many initial guesses and obtained the set of unique solutions. Specifically, the photon and carrier number guesses were simply set to the free-running values and a range of values for the phase offset ($\tilde{\phi}$) between 0 and $2\pi$ and frequency ($\omega$) values ranging from --10 to 10 GHz. We found that the locked frequency is mainly what differentiated the solutions so by default our code tested 20 phase offset values and 200 frequencies at each phase offset. More guesses can be intelligently added as shown in Sec.~\ref{sec:results} to obtain more continuous branches of equilibrium solutions.

\subsection{Linearization}

Using the equilibrium information from Sec.~\ref{sec:eq_comp}, we aim to find a linear system centered at the origin that approximates the full system behavior near the equilibrium point in the form
\begin{equation}\label{eq:linear_dde}
    \mathbf{\dot{x}}=\mathbf{A}_0\mathbf{x}(t)+\mathbf{A}_1\mathbf{x}(t-\tau),
\end{equation}
where $\mathbf{x}$ corresponds to the original states of Eq.~\ref{eq:nd_comp_model} and $\mathbf{A}_m\in\mathbb{R}^{3N\times 3N}$ is the Jacobian matrix of Eq.~\ref{eq:nd_comp_model} evaluated at $\mathbf{x}^*$ corresponding to $\mathbf{x}(t-m\tau)$ for $m\in[0,1]$
\begin{equation}
    \mathbf{A}_m =
\begin{bmatrix}
\mathbf{A}_m^{(1,1)} & \mathbf{A}_m^{(1,2)} & \dots & \mathbf{A}_m^{(1,N)} \\
\mathbf{A}_m^{(2,1)} & \mathbf{A}_m^{(2,2)} & \dots & \mathbf{A}_m^{(2,N)} \\
\vdots & \vdots & \ddots & \vdots \\
\mathbf{A}_m^{(N,1)} & \mathbf{A}_m^{(N,2)} & \dots & \mathbf{A}_m^{(N,N)}
\end{bmatrix},
\end{equation}
where each element of $\mathbf{A}_m$ consists of a $3\times 3$ block matrix of the form
\begin{alignat}{2}
\added{\mathbf{A}_0^{(i,j)}} &=
\begin{pmatrix}
\frac{\partial\dot{n}_i}{\partial n_j} &
\frac{\partial\dot{n}_i}{\partial s_j} &
\frac{\partial\dot{n}_i}{\partial \phi_j} \\
\frac{\partial\dot{s}_i}{\partial n_j} &
\frac{\partial\dot{s}_i}{\partial s_j} &
\frac{\partial\dot{s}_i}{\partial \phi_j} \\
\frac{\partial\dot{\phi}_i}{\partial n_j} &
\frac{\partial\dot{\phi}_i}{\partial s_j} &
\frac{\partial\dot{\phi}_i}{\partial \phi_j}
\end{pmatrix}
\quad &
\added{\mathbf{A}_1^{(i,j)}} &=
\begin{pmatrix}
\frac{\partial\dot{n}_i}{\partial n_j^\tau} &
\frac{\partial\dot{n}_i}{\partial s_j^\tau} &
\frac{\partial\dot{n}_i}{\partial \phi_j^\tau} \\
\frac{\partial\dot{s}_i}{\partial n_j^\tau} &
\frac{\partial\dot{s}_i}{\partial s_j^\tau} &
\frac{\partial\dot{s}_i}{\partial \phi_j^\tau} \\
\frac{\partial\dot{\phi}_i}{\partial n_j^\tau} &
\frac{\partial\dot{\phi}_i}{\partial s_j^\tau} &
\frac{\partial\dot{\phi}_i}{\partial \phi_j^\tau}.
\end{pmatrix}
\end{alignat}
In terms of system parameters from Eq.~\eqref{eq:nd_comp_model}, the Jacobian matrices evaluated at $\mathbf{x}^*$ are expressed as,
\begin{equation}
\begin{aligned}
    \mathbf{A}_0^{(i,i)} &=
\begin{pmatrix}
-\frac{1}{T}\left(1 + \frac{s_i^*}{(1+\added{\gamma'} s_i^*)}\right) & -\frac{1}{T}\left(\frac{(1+n_i^*)(1+\added{\gamma'} s_i^*-\added{\gamma'})}{(1+\added{\gamma'} s_i^*)^2}\right) & 0 \\
\frac{s_i^*}{1+\added{\gamma'} s_i^*} + \beta & \frac{(1+n_i^*)(1+\added{\gamma'} s_i^*-\added{\gamma'})}{(1+\added{\gamma'} s_i^*)^2}-1 + \sum\limits_{j\neq i} \kappa'_{ij} \sqrt{\frac{s_j^*}{s_i^*}}\mathcal{C}(\tilde{\phi}_{ij}^p)  & 2 \sum\limits_{j\neq i} \kappa'_{ij} \sqrt{s_i^* s_j^*} \mathcal{S}(\tilde{\phi}_{ij}^p) \\
\frac{\alpha}{2(1+\added{\gamma'} s_i^*)} & - \frac{\alpha}{2} \frac{(n_i-\bar{n}) \added{\gamma'}}{(1+\added{\gamma'} s_i^*)^2} - \frac{1}{2} \sum\limits_{j\neq i} \kappa'_{ij} \sqrt{\frac{s_j^*}{{s_i^*}^{3}}}\mathcal{S}({\tilde{\phi}_{ij}^p}) & - \sum\limits_{j\neq i} \kappa'_{ij} \sqrt{\frac{s_j^*}{s_i^*}}\mathcal{C}(\tilde{\phi}_{ij}^p) 
\end{pmatrix},\\
\mathbf{A}_0^{(i,j)} &=
\begin{pmatrix}
0 & 0 & 0 \\
0 & 0 & 0 \\
0 & 0 & 0 \\
\end{pmatrix},
\end{aligned}
\end{equation}
where $\mathcal{S}(\cdot)\equiv\sin{(\cdot)}$, $\mathcal{C}(\cdot)\equiv\cos{(\cdot)}$, and $\tilde{\phi}_{ij}^p = \phi_j^*-\phi_i^*-\omega\tau-\phi_p$. Because all subscript $j$ terms depend on the time delay, all $\mathbf{A}_0^{(i,j)}$ are $3\times 3$ zero matrices. Similarly, we also compute the elements of $\mathbf{A}_1$ as,
\begin{equation}
\begin{aligned}
\mathbf{A}_1^{(i,i)} &=
\begin{pmatrix}
0 & 0 & 0 \\
0 & 0 & 0 \\
0 & 0 & 0 \\
\end{pmatrix},\\
\mathbf{A}_1^{(i,j)} &=
\begin{pmatrix}
0 & 0 & 0 \\
0 & \sum\limits_{j\neq i} \kappa'_{ij} \sqrt{\frac{s_i^*}{ s_j^*}} \mathcal{C}(\tilde{\phi}_{ij}^p) & -2 \sum\limits_{j\neq i} \kappa'_{ij} \sqrt{s_i^* s_j^*} \mathcal{S}(\tilde{\phi}_{ij}^p) \\
0 & \frac{1}{2}\sum\limits_{j\neq i} \kappa'_{ij} \frac{1}{\sqrt{s_i^* s_j^*}} \mathcal{S}(\tilde{\phi}_{ij}^p) &  \sum\limits_{j\neq i} \kappa'_{ij} \sqrt{\frac{s_j^*}{ s_i^*}} \mathcal{C}(\tilde{\phi}_{ij}^p)\\
\end{pmatrix}.
\end{aligned}
\end{equation}

\subsection{Stability}
To determine the stability of $\mathbf{x}^*$ using the linearized model [Eq.~\eqref{eq:linear_dde}], we assume a complex exponential solution $\mathbf{x}=\exp{(\lambda t)}$, where $\lambda$ is a diagonal matrix of eigenvalues that characterize the stability of $\mathbf{x}^*$. Substituting this assumed solution into Eq.~\eqref{eq:linear_dde}, we get
\begin{equation*}
\lambda\exp{(\lambda t)}=\mathbf{A}_0\exp{(\lambda t)} + \mathbf{A}_1 \exp{(\lambda (t-\tau))}.
\end{equation*}
Because $\exp{(\lambda t)}>0~\forall t$, we divide through by this term to get
\begin{equation*}
\lambda\mathbf{I}-\mathbf{A}_0 - \mathbf{A}_1 \exp{(-\lambda \tau)}=0.
\end{equation*}
Solving this equation for nontrivial $\lambda$ provides stability information about $\mathbf{x}^*$. Specifically, if any $\Re{(\lambda})>0$, we say the equilibrium point is unstable. Nontrivial $\lambda$ solutions occur when
\begin{equation}\label{eq:char_eq}
\det{(\lambda\mathbf{I}-\mathbf{A}_0 - \mathbf{A}_1 \exp{(-\lambda \tau)}})=0.
\end{equation}
Equation~\eqref{eq:char_eq} is the characteristic equation for the system. We remark that this equation is transcendental due to the time delay terms and is satisfied by an infinite spectrum of eigenvalues $\lambda$. While in practice it is impossible to solve for all eigenvalues of Eq.~\eqref{eq:char_eq}, we only need the right most $\lambda$ in the complex plane to fully characterize the stability of $\mathbf{x}^*$ and it has been proven that for linear DDEs the number of eigenvalues with $\Re{(\lambda})>0$ is finite \cite{smith_linear_2011,stepan_rds_1989} due to the exponential decay as $\lambda\to\infty$. Solving Eq.~\eqref{eq:char_eq} numerically like we did for finding the equilibrium points of the system is computationally expensive due to the determinant and the need to guess different eigenvalues so the solver can converge on a true solution. This problem is further compounded by the high dimensionality of the system and for large laser arrays the problem quickly becomes intractable. To avoid this issue, we turned to spectral discretization methods such as Chebychev collocation. This approach works by discretizing the history function at carefully chosen points to obtain a discrete approximation for the infinite dimensional solution operator. We refer the reader to Refs.~\cite{wu_reliably_2012, michiels_stability_2014, jarlebring_krylov_2010, breda_pseudospectral_2005} for more information on how this method works. This approach was well matched to our setting, where the goal was to analyze the spectra of many precomputed equilibria in heterogeneous delay-coupled arrays rather than to continue individual branches. The collocation method produced a direct matrix eigenvalue problem at each equilibrium. While \texttt{DDE-BIFTOOL} \cite{noauthor_dde-biftool_nodate} is a commonly used, powerful package for continuation and local bifurcation analysis of delay differential equations, its workflow is centered on branch continuation and separate computation of the rightmost characteristic roots. Due to the large set of coexisting equilibria considered here, the collocation approach provided a more direct and reliable route to stability classification.

\makeatletter
\immediate\write\@auxout{\string\citation{apsrev42Control}}
\makeatother
\bibliographystyle{apsrev4-2-urls}
\bibliography{aps-control,references, manual_refs}

\end{document}